\documentclass[prd,tightenlines,
nofootinbib,amsfonts,amssymb,amsmath,color,11pt]{revtex4}

\usepackage{color}
\usepackage{graphicx}
\usepackage{dcolumn}
\usepackage{bm}

\newcommand\mpl{M_{Pl}}

\def\be{\begin{equation}} 
\def\ee{\end{equation}}
\def\bea{\begin{eqnarray}}
\def\eea{\end{eqnarray}}

\begin{document}
	
	\title[]{Compact objects in scalar-tensor theories after GW170817}

	\author{Javier Chagoya}\email{jfchagoya@gmail.com}
	\author{Gianmassimo Tasinato}%
\email{g.tasinato2208@gmail.com}
	\affiliation{ 
	Department of Physics, Swansea University, Swansea, SA2 8PP, U.K.
	}%

\begin{abstract}
The recent observations of neutron star mergers have changed our perspective on scalar-tensor theories of gravity,
favouring models where gravitational  waves travel at the speed of light. In this work we consider a scalar-tensor set-up
with such a property, belonging to a beyond Horndeski system, and we numerically investigate the physics of locally asymptotically 
flat black holes and relativistic stars. We first determine  regular black hole solutions equipped with horizons: they are 
characterized by a deficit angle at infinity, and by large contributions of the scalar to the geometry in  the near horizon region. 
 We then study  configurations  of incompressible relativistic  stars.  We show 
  that their compactness can be much higher than  stars with the same energy density in  General Relativity, and
   the scalar field profile 
    imposes stringent    constraints on the star properties.
    These results can suggest new ways
  to probe the efficiency of  screening mechanisms in strong gravity  regimes, and can help to build specific observational tests for scalar-tensor gravity models 
  with unit 
  speed for gravitational waves.  
\end{abstract}
	
	\maketitle
	
	\section{Introduction}
	
	Scalar-tensor theories with non-minimal couplings  between
        scalar fields and gravity find interesting 
        applications to  cosmology (dark energy and dark
        matter problems, see e.g. the review \cite{Clifton:2011jh}) and  quantum gravity (including Lorentz violating systems as
        Horava-Lisfshitz gravity, see \cite{Wang:2017brl} for a recent review). Moreover, they  are able to screen fifth forces
        by means of the Vainshtein mechanism (see for example \cite{Amendola:1993uh,Kimura:2011dc,Babichev:2013usa,Bellini:2015wfa}).
 Over the years, many advances
  have been made in developing  consistent scalar-tensor theories, going from Brans-Dicke systems \cite{Brans:1961sx},
        to Galileons and Horndeski theories \cite{Nicolis:2008in, Horndeski:1974wa}, to beyond Horndenski and
        DHOST/EST scenarios \cite{	Gleyzes:2014dya,Zumalacarregui:2013pma, Langlois:2015cwa, Crisostomi:2016czh, BenAchour:2016fzp, Crisostomi:2016tcp}.  
        The study of black holes and compact
        relativistic stars in these richer scalar-tensor theories is relevant for
        phenomenological investigations
          of screening mechanisms inside compact sources \cite{Herdeiro:2015waa, Sakstein:2015aac, Silva:2016smx, Sakstein:2016oel}, and for our theoretical
        understanding of no-hair and singularity theorems in Einstein General
        Relativity (GR) non-minimally coupled with    scalar fields 
 (see e.g. the discussion in       \cite{Bekenstein:1996pn}).
         The purpose of this work is to investigate asymptotically flat  black holes and relativistic stars in a class of scalar-tensor theories compatible
         with the stringent constrains recently obtained from the observation of neutron star mergers. 
        
         Asymptotically flat black hole
        solutions with non-trivial scalar  profiles have been found in
        Horndeski gravity (see \cite{Babichev:2016rlq} for a review), and some are known
        for beyond Horndeski theories     \cite{Babichev:2017guv}. A non-vanishing scalar field profile
         may or may not affect the properties of the geometry. Even if  black hole solutions in these theories
        exhibit only 
        small deviations from their GR counterparts, it is  possible  
        that scalar field effects become more relevant in presence of matter,
        thus leading to sizeable consequences  that can   be constrained with
        observational data. This phenomenon was first pointed out in a theory of Brans-Dicke
        gravity applied to neutron star objects in  \cite{Damour:1993hw, Damour:1996ke, Salgado:1998sg} and dubbed  \emph{spontaneous scalarisation}; it
        has also  been analysed recently in more general scalar-tensor theories
        \cite{Chagoya:2014fza, Kobayashi:2014ida,Crisostomi:2017lbg,Dima:2017pwp,Silva:2017uqg}.
        Investigations of explicit solutions for compact relativistic objects  are necessary for acquiring    a
        better understanding of how these systems can be distinguished  from GR. Studies along these
        lines typically focus on neutron stars, since the     
        strong gravitational field around these objects provides a good
        laboratory to test modified gravity theories. These investigations   have shown
        that configurations compatible -- within the error bars -- with the measured  masses and radii  of
        neutron stars are common      in scalar-tensor gravity \cite{Cisterna:2015yla, Maselli:2016gxk,Babichev:2016jom}. The
        analysis of these systems in modified gravity is  at an early stage
        in comparison with the theoretical advances made in GR over the past decades, although new developments
        concerning equation-of-state independent relations between
        properties of relativistic compact objects indicate promising  tools to constrain modified gravity theories
        \cite{Pani:2014jra,Pappas:2014gca,Gupta:2017vsl}.

       Recently, gravitational and electromagnetic radiation emitted by  NS mergers 
       was detected almost simultaneously by LIGO, VIRGO, and an array of 
       observatories on earth and in space \cite{TheLIGOScientific:2017qsa}, placing  a strong constraint on the difference 
       between the propagation speed of gravitational waves ($c_{GW}$) and the speed of light,
       $-3\times 10^{-15} \leq c_{GW} - 1  \leq 7\times10^{-16}$ \cite{Monitor:2017mdv}, where the speed of light is normalized to unity.  
       Besides  the quadratic and cubic Horndeski Lagrangians, scalar-tensor theories 
       of gravity generically predict gravitational waves that do not travel at the speed of light. 
       There are, on the other hand,  particular combinations of Horndeski and beyond Horndeski Lagrangians which  
      predict $c_{GW} = 1$ \cite{Ezquiaga:2017ekz,Creminelli:2017sry,Sakstein:2017xjx,Baker:2017hug}. 
      Observational consequences  of these 
       Lagrangians have been recently analysed, and it has 
       been shown that  -- in absence of a canonical kinetic term for a scalar field --  the screening mechanism 
       allows to recover exact GR solutions in vacuum, although screening effects are  broken in presence of matter 
       \cite{Crisostomi:2017lbg,Dima:2017pwp}.
	  
	      
	      On the other hand, on general physical grounds we expect that the standard scalar kinetic term should be present in the scalar action, being the leading
	      dimension four operator that governs the scalar dynamics, at least around  nearly flat backgrounds. 
	      In this paper we focus on a specific  scalar-tensor theory that includes, besides the kinetic term of the scalar field,  
	      a combination of quartic Horndeski and beyond Horndeski contributions  satisfying the condition $c_{GW} = 1$. The presence of the standard kinetic terms
	      affects considerably the geometry, and we find several new phenomena associated with the non-linearity of our system of equations. 
	       In Section~\ref{sec:model} we present the theory under consideration.
	  In Section~\ref{sec:bhsol} we determine the conditions to satisfy for obtaining (locally) asymptotically flat  
	      black hole solutions  supporting a non-trivial scalar field profile.  We
	      numerically analyse
	       how the scalar  affects the size of the horizons, and we find conditions to avoid naked singularities. The corresponding 
	      solutions are characterized by  a deficit angle induced by the scalar field
	      kinetic terms. In Section~\ref{sec:kin4mat} we proceed to study relativistic compact objects in this theory
	      and we find that, in contrast to other scenarios of beyond Horndeski systems, the angular deficit does not produce a singularity at the centre of these objects.
	      The non-linearities of the equations lead to new phenomenological consequences, as {for example specific 
	       relations between radius and energy content
	      of the objects we investigate.}
	       By matching interior and exterior solutions we find situations where the
	      scalar field contributions to the geometry are dominant inside a compact object, but negligible  in the exterior, pointing towards 
	      a sizeable  breaking of a Vainshtein screening mechanism. Finally, we discuss how the compactness of this scalar-tensor
	      configurations, and the  deficit angle itself, can be used to constrain this theory.
	    
	\section{Theories of Horndeski and beyond after GW170817}\label{sec:model}
	
		The most general scalar-tensor theory leading to second order equations of motion is a combination of the Horndenski Lagrangians \cite{Horndeski:1974wa}. Calling $\phi$
	the scalar field, such Lagrangian densities are
	\cite{Deffayet:2011gz, Kobayashi:2011nu}
	\begin{align}
		\mathcal L_2 & = G_2, \hspace{3em} \nonumber \\ 
		\mathcal L_3 & = G_3 \,  [\Phi] \, , \nonumber \\
		\mathcal L_4 & = G_4 R + G_{4,X} \left\{[\Phi]^2 - [\Phi^2]   \right\},
		\nonumber\\
		\mathcal L_5 & = G_5 G_{\mu\nu} \Phi^{\mu\nu} - \frac{1}{6}G_{5,X}
		\left\{[\Phi]^3 
		- 3 [\Phi][\Phi^2]+ 2 [\Phi^3] \right\}, \label{eq:hlag}
	\end{align}
where {$\Phi$ is a matrix with components $\nabla^\mu\nabla_\nu \phi$ }, and
\begin{align}
X =&-\frac12\,\partial_\mu\phi \partial^\mu \phi \, ,
\nonumber
\\
	 \left[ \Phi^n \right]  = & \, {\text{tr}} \left( \Phi^n \right) \, ,
	 \nonumber\\
	 \langle \Phi \rangle =& \, \partial^\mu \phi \partial_\mu\partial_\nu \phi \partial^\nu \phi\,.
\end{align}
	   $G_i$ are arbitrary functions
	of $\phi$ and $X$, or only of $X$  if we impose a shift symmetry $\phi\to\phi+$const.  The equations
	of motion associated with Lagrangians~(\ref{eq:hlag}) are second order
	 ensuring   that the system is free of Ostrogradsky instabilities. On the
	the other hand, it is possible to have healthy   scalar-tensor theories 	also with higher order equations of motion, provided that constraint conditions
	 forbid the propagation of would-be Ostrogradsky ghosts. Explicit examples are the theories of 
	 beyond Horndeski \cite{Gleyzes:2014dya}, and their generalizations dubbed DHOST/EST theories \cite{Langlois:2015cwa,Crisostomi:2016czh}.  The theory of 
	 beyond Horndeski  is constructed with the  Lagrangian densities
	  \begin{align}
		\mathcal L_4^{bH} & = -\frac{1}{2} F_4\, \epsilon^{\mu\nu\rho}{}_{\sigma}
		\epsilon^{\mu'\nu'\rho'\sigma} \partial_\mu \phi \partial_{\mu'} \phi
		\Phi_{\mu\mu'}\Phi_{\nu\nu'}\Phi_{\rho\rho'}\, , \nonumber \\
		\mathcal L_5^{bH} & = F_5\,
		\epsilon^{\mu\nu\rho\sigma}\epsilon^{\mu'\nu'\rho'\sigma'} \partial_\mu \phi
		\partial_\mu' \phi \Phi_{\nu\nu'}\Phi_{\rho\rho'}\Phi_{\sigma\sigma'}\, ,
	\end{align}
with $F_{4,\,5}$ arbitrary functions of $\phi$, $X$. 
	 Theories described by a combination of the previous Lagrangians -- apart from systems only containing $\mathcal L_2$ and $\mathcal L_3$ --
	 generally lead to a modification of the speed of propagation of gravity waves, hence they are disfavoured by the recent observation of gravitational waves from a neutron star merger 
	 GW170817 and its associated electromagnetic counterpart GRB 170817A. On the other hand, there are  specific combinations of the Horndeski and beyond
	  Horndeski      Langrangians which do not change the speed of gravitational waves \cite{Bettoni:2016mij}. A particular example is the combination 
	 	\begin{align}
	\mathcal L_c & = X +  \mathcal L_4 + \mathcal L_4^{bH}, \hspace{1em}\text{with }
	F_4 = G_{4,X}/X\, , \nonumber \\
	& = X + G_4 R + \frac{G_{4,X}}{X} \left(\langle\Phi^2 \rangle -
	\langle\Phi\rangle [\Phi]   \right)\, , \label{ac:lc}
	\end{align}
which we consider in this work.  This Lagrangian density includes the standard scalar kinetic term, accompanied by derivative self-interactions and non-minimal
couplings with the metric  which become important in strong gravity regimes such as in proximity of black holes or in dense objects. For simplicity, and definiteness,
we study this theory with the function $G_4$ chosen as
\be
G_4 = M_{Pl}^2 + \frac{\beta}{M_{Pl}^{2}} \, X \label{eq:g4choice}\, ,
\ee 
	where $\beta$ is a dimensionless constant.
 {Black hole configurations for similar systems have been studied in the past 
 	both in beyond Horndeski and vector-tensor theories \cite{Gripaios:2004ms,Tasinato:2014eka,Heisenberg:2014rta}. In
	particular, a stealth Schwarzschild solution was fist discovered  for vector-tensor systems with the same choice of $G_4$ and
	a special value of $\beta$ \cite{Chagoya:2016aar}. When the time component $A_0$ of the vector is constant, this solution is equivalent to  a scalar-tensor
	stealth configuration for a scalar field of the form $\phi \,= \,q \,t + \phi_1(r)$, with a constant $q$ \cite{Babichev:2013cya}. Further generalisations based on this solution can be found in
	\cite{Heisenberg:2017hwb,Chagoya:2017fyl}, where neutron stars and asymptotically flat black holes are constructed
	for arbitrary values of $\beta$ and vector-tensor  generalisations of~(\ref{eq:g4choice}). 
	}

	In this work we focus on the scalar-tensor theory \eqref{ac:lc}, studying new black hole solutions in vacuum
	with novel features (Section~\ref{sec:bhsol}), and the physics of gravitationally bound compact  objects made of incompressible matter (Section~\ref{sec:kin4mat}).

	\section{Black holes }\label{sec:bhsol}

	The study of black hole solutions in vacuum for scalar-tensor theories with non-minimal couplings to gravity is interesting at least for two reasons.  First, it allows 
	 to probe a strong gravity regime for the theory one considers, where  non-perturbative 
	  contributions to screening mechanisms can make manifest sizeable deviations
	  from GR results (see, e.g.  \cite{Saito:2015fza, Koyama:2015oma}).
	  Second, it allows to test no-hair and singularity theorems  in new settings, possibly revealing
	   new geometries or topologies characterized  by additional scalar  charges (see, e.g.  \cite{Achucarro:1995nu,Bekenstein:1996pn}).
	    In this section  we aim to investigate whether there exist    asymptotically flat black hole configurations for the beyond Horndeski theory
	    of Lagrangian \eqref{ac:lc}, answering almost affirmatively -- in the sense that we
 find  \emph{locally}  asymptotically flat black hole solutions, for which the curvature invariants vanish for large $r$, but that are characterized
 by a constant angular deficit at infinity. 
 The existence of an angular deficit in beyond Horndeski theories was first identified in \cite{DeFelice:2015sya} as a potential source of
 singularities at the centre of configurations of matter. As we show below, both in vacuum and inside compact objects
 the angular deficit of the model we are considering does not affect the regularity of the solutions, provided that some conditions are satisfied.
 
	{The covariant form of the equations of motion (EOMs) for the the scalar $\phi$ and the metric $g_{\mu\nu}$ is given in Appendix~\ref{app:a}.} Since we are interested on static, spherically symmetric space-times, we start imposing  the following Ansatz for the metric 		
			\begin{equation}
	ds^2 = -f(r) dr^2 + h(r)^{-1} dr^2 + r^2 d\theta^2 + r^2 \sin^2\theta
	d\varphi^2\, , \label{eq:ansatzmetricnodeficit}
	\end{equation}
while we allow for a linear time dependence in the scalar configuration  
\begin{equation} \label{scal-ans1}
\phi =  \mpl^2 \phi_0\,	t + \phi_1(r) \, ,
\end{equation}
where $\phi_0$ is a dimensionless constant\footnote{From now on we set $\mpl=1$. The correct dimensions of 
	all expressions are recovered after reinstating the appropriate factors of $\mpl$.}. This Ansatz for the scalar field  is compatible with a static spacetime 
	 (recall that the equations of motion always contain derivatives of the scalar) 
	and have been extensively
 studied  in the recent literature on scalar-tensor black hole solutions, 
 since the time dependence explicitly breaks the assumptions of no-hair theorems in Horndeski theory \cite{Babichev:2016rlq}, thus opening up the possibility of finding asymptotically flat black holes dressed with a scalar field (see, e.g. \cite{Babichev:2013cya, Babichev:2016rlq, Babichev:2017guv}, and the review
 \cite{Herdeiro:2015waa}). 

 Using  these  Ansatz for metric and scalar, we find that the $(t,r)$ component of the metric EOMs (the $\xi_{tr}$ component in eq \eqref{eq:meteq}) reduces to an algebraic condition for the derivative of the radial scalar field profile $\phi_1$,
 which reads
 
 \begin{equation}
	\frac{\beta  \phi' _1 \left[f^2 h (h+1) \phi' _1{}^2+f \left(h^2 r \phi'
		_1{}^2 f'+ \phi _0^2 \left(r h'-1\right)+h  \phi _0^2\right)-2 h r \phi _0^2
		f'\right]}{ f r^2 \left(f h \phi' _1{}^2-\ \phi _0^2\right)}+\frac{1
	}{2} \phi' _1\, = \,0\, .
	\end{equation}

If one chooses $\beta\,=0$ --
 corresponding to GR plus a standard kinetic term for the scalar field --  the only solution of the previous equation is $\phi' _1 \,= \,0$. On the other hand, if $\beta\neq0$, we have a cubic equation for $\phi_1'$, which additionally
admits the following two branches of solutions: 

\begin{equation} \label{phi1ps}
	\phi_1' = \pm \phi _0 \sqrt{ \frac{ 4 \beta  h r f'+  f 
			h^2 r^2+2  \beta  f h^2+2 \beta  f r h'-2 \beta  f h}{  2\beta  f^2 + f^2 
			h r^2 + 2 \beta  f^2 h +2 \beta  f r f' }}\, .
	\end{equation}

Notice that such branches are well defined also in the limit $\beta\to0$, giving $\phi_1' = \pm  \phi _0$: hence these branches are disconnected from the   
$\beta=0$ branch, even  when $\phi_0$ is turned on.  The
	presence of different branches is common in Horndeski and beyond Horndeski theories
	where the scalar field derivative  satisfies a non-linear algebraic equation,
	and the non-trivial scalar field profile is responsible for providing a screening
	mechanism that  recovers GR solutions in the strong gravity regime \cite{Babichev:2013usa, Babichev:2016jom, Crisostomi:2017lbg, Dima:2017pwp}.
	 In what follows, we will concentrate on the upper branch of the algebraic solution \eqref{phi1ps}. 
 The remaining independent equations, that we take as  the $(t,t)$ and $(r,r)$ components of the metric equations, 
 %
 are hard to solve exactly for $f, \,h$, but we can study the system numerically, or analytically in certain regimes.
 
Flat space, corresponding to the choice $f\,=\,h\,=\,1$ using Ansatz \eqref{eq:ansatzmetricnodeficit},  is {\it not} a solution of the EOMs.  
Asymptotically de Sitter solutions can be easily found (very similar to the ones originally found in
 \cite{Rinaldi:2012vy}),  but here we will focus on black hole solutions that are at least locally 
asymptotically flat. This branch of solutions has been less studied
in the literature,  and it is important to investigate in detail the corresponding phenomenology.  
In order to take into account local instead of global flatness, 
it is compulsory to slightly generalize the metric Ansatz~(\ref{eq:ansatzmetricnodeficit}) by including
a deficit angle,
\begin{equation}
	ds^2 = -f(r) dt^2 + h(r)^{-1} dr^2 + s_0^{-1} \,r^2 d\Omega^2 \, ,
	\label{eq:metansatzwiths}
	\end{equation} 
with $s_0$ not necessarily equal to one. This  modification does not change the branch structure of the solutions of the scalar field equation.
 With such Ansatz, it is possible to analytically determine  asymptotic solutions for the functions $f,\,h$
expanded in inverse powers of the radial distance $r$, imposing the condition that $f\,=\,h\,=\,1$ at asymptotic infinity. 
The corresponding equations of motion with this Ansatz are given in~(\ref{eq:ssstt}-\ref{eq:ssstr}).
We find, up to second order in an $1/r$ expansion,  
	\begin{align}
	s_0 & = {1-3 \beta \phi _0^2}\, , \\
	f(r) & = 1 - \frac{2M}{r} -\frac{4 \beta ^2 \phi _0^2 \left(\beta \phi _0^2-2
		\right)}{  r^2}+ {\cal O}\left(\frac{1}{r^3}\right)\, ,  \label{eqs:asympkinsolf}\\
	h(r) & = 1 - \frac{2M}{r} + \frac{4 \beta ^2 \phi _0^2
		\left(1-\beta  \phi _0^2\right)}{r^2}+ {\cal O}\left(\frac{1}{r^3}\right)\,\, ,
	\label{eqs:asympkinsolh} \\
	\phi'_1(r) & = \phi _0 +\frac{2 M  \phi _0}{r} +  \frac{2  \phi _0 \left[2 \beta ^3
		\phi _0^4+ \left( 2 M^2  - \beta \right)-3 \beta ^2  \phi _0^2
		\right]}{ r^2 } + {\cal O}\left(\frac{1}{r^3}\right)\,\, ,
	\label{eqs:asympkinsolphi}
	\end{align}
with $M$ an integration constant. 
We notice that, since $s_0\neq1$, the geometry has a deficit angle: this  is a consequence of including the kinetic
terms of the scalar in our action. On the other hand,  the radial dependence
of the functions $f$, $h$ gives us   hope  that a would-be conical singularity at the origin  $r=0$ can be 
absent, or covered by   horizons.  In what follows, we discuss  conditions for ensuring  that this is the case for
 the system under consideration.  Notice that, besides the deficit angle, the standard `$1-2 M/r$' behaviour (plus subleading corrections)
 of the metric components indicates that the metric is asymptotically flat and approaches  GR results at  large distances. 
 
 Conical deficits covered by horizons  have a long history in black hole physics, starting from \cite{Aryal:1986sz},  and
  physical realizations  and interpretations --  related with  strings piercing the
 black hole horizons in Abelian-Higgs models  \cite{Achucarro:1995nu} --  can be subtle \cite{Bekenstein:1996pn}.
  It is interesting that conical deficits  appear also in the context of a single scalar field coupled with gravity, and we will
 later attempt to connect them  with no hair theorems for this system.   Geometries with similar deficit angles arise
  when considering gravitational monopoles  \cite{Barriola:1989hx,Shi:2009nz,Nucamendi:1996ac}. 
 
 The presence of conical singularities
 in  solutions of beyond Horndeski theories has been first pointed out in \cite{DeFelice:2015sya,Kase:2015gxi}: they focus on systems that are not shift symmetric, finding harmful  conical singularities at the origin unless the parameters of the theory are appropriately tuned. A set-up more similar to ours has been analysed in a  vector-tensor system \cite{Heisenberg:2016lux}, showing that conical singularities can then be avoided. We will make more detailed comparisons with these works in later
 Sections.
 
  \subsection{Numerical evidence for regular black holes}
  We now  provide numerical evidence that spherically symmetric, locally asymptotically flat solutions  of the EOMs~(\ref{eq:ssstt}-\ref{eq:ssstr})
  are free of  naked conical singularities at the origin,  when appropriate  conditions are satisfied.  As we shall see, despite the fact that the solution for 
  the metric components has the standard  $1/r$ behaviour at large distances from the origin, there
  arise large deviations from GR configurations 
  near the black hole horizon. 
  
\begin{figure}
		\includegraphics[height=5.4cm]{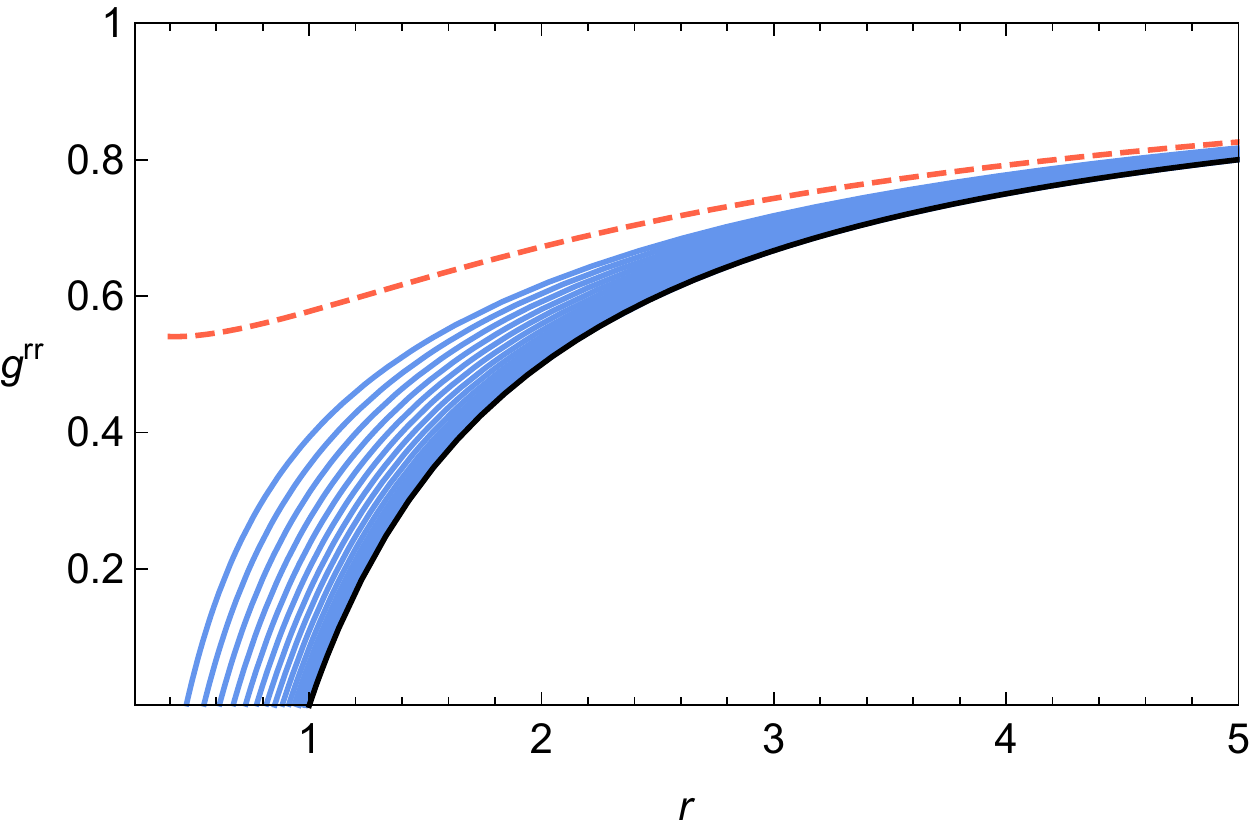} \ \includegraphics[height=5.4cm]{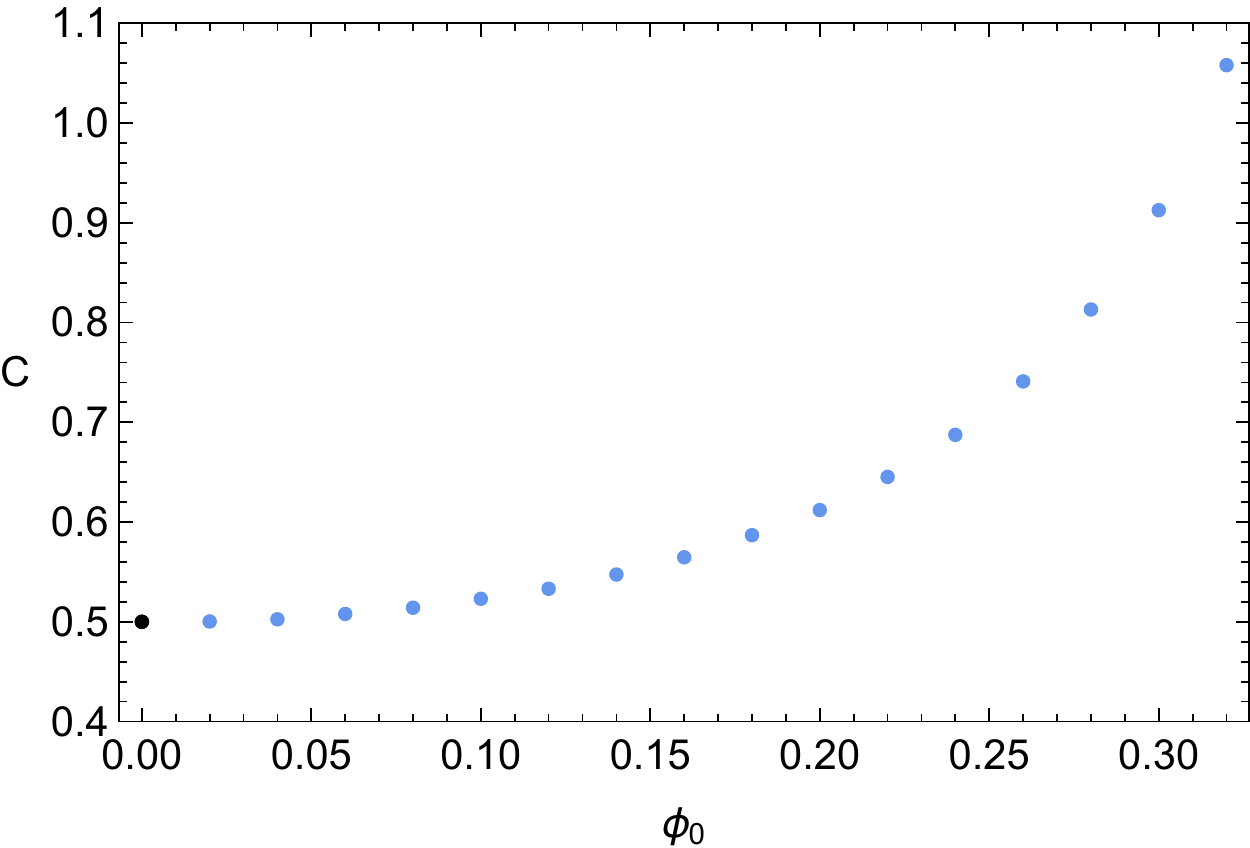}
		\caption{\it{Numerical BH solutions for $2  M = \beta = 1 $. The left panel shows the metric component $g_{rr}$ in GR (black line) and in the theory we consider,  with $\phi_0$ spanning 
		between $0.02$ and $0.32$ (blue lines). The size of the black hole horizon decreases as $\phi_0$ increases. For $\phi_0\gtrsim 0.40$ we do not find solutions that form an event horizon. The dashed line shows a solution for $\phi_0 = 0.41$. 
			The compactness of these black holes is shown in the right panel. } }
		\label{fig:bhsols}
	\end{figure}

  We numerically solve equations~(\ref{eq:ssstt}-\ref{eq:ssstr}),  using the asymptotic fields given in 
	eqs.~(\ref{eqs:asympkinsolf}) and (\ref{eqs:asympkinsolh}) 
	as boundary conditions, and  proceed integrating inwards towards small $r$ until we encounter the position $r_{h}$ of a horizon,  defined by the condition
	$g^{rr} = h(r_{h}) = 0$. 
	The system of equations~(\ref{eq:ssstt}-\ref{eq:ssstr}) is reduced to two
	equations for $f$ and $h$, since we  impose 
	$s_0 = 1 - 3 \beta \phi_0^2$ and we algebraically solve the equation 
	$\xi_{tr}\,=\,0$  for $\phi_1'$.
	We fix $\beta = 1$, so that the size of the angular deficit is controlled
	only by the scalar  parameter $\phi_0$.  The boundary conditions are then specified in terms of two quantities: the black 
	hole mass $M$, and $\phi_0$. For definiteness we fix $M=0.5$  (recall that we work in units where $M_{Pl}=1$)     and construct black hole solutions characterised by   different  values of $\phi_0$. In order to ensure that the conical deficit is positive ($s_0>0$ in eq~(\ref{eq:metansatzwiths})), we limit our
	investigation to the interval $0\le \phi_0  < 1/\sqrt{3 \beta} \sim 0.57$.
	 Our numerical results are  shown in Fig.~\ref{fig:bhsols}.

	The left panel of Fig.~\ref{fig:bhsols}  shows $h(r)$ -- the inverse of
	the radial metric 
	component. The black line shows the quantity $h(r)$ for the Schwarzschild metric, and the blue
	lines correspond to
	different values of $\phi_0$ between $0.02$ and $0.32$.   
	For each of the blue lines
	 the function $h(r)$ crosses zero, indicating the position of  an horizon, whose size 
	shrinks as $\phi_0$ increases. Solutions for $0.32<\phi_0 \lesssim 0.40$ can be found as well, but they require a higher numerical precision near the horizon. For $\phi_0 \gtrsim 0.40$ we do not find regular solutions equipped with an horizon: an interpretation for this fact will be  provided below.
	
	 The right panel shows the compactness of
	such black holes,
	\begin{equation}
	C_{BH} = \frac{M}{r_h},
	\end{equation}
	where $r_h$ is the radius of the event horizon, and $M=0.5$ is fixed by means
	of 
	the asymptotic conditions (\ref{eqs:asympkinsolf}) and (\ref{eqs:asympkinsolh}).
	The point in black is the compactness of the Schwarzschild black hole, $C_{Schw}
	= 0.5$.
	The compactness increases non-linearly with  $\phi_0$, showing that -- thanks to the ${\cal O}(1/r^2)$ corrections to the metric -- 
	our solutions are different from the Schwarzschild configuration when
	approaching the horizon. 
	
	\smallskip
	Let us return to  specifically discuss the behaviour of the system for $\phi_0 \gtrsim 0.40$.
	 For our
	values of $M=0.5$ and $\beta=1$, we could not find solutions with an event horizon
	for $\phi_0 \gtrsim 0.40$. Indeed, when changing from $\phi_0 = 0.40$ to
	$\phi_0 = 0.41$ the solution for the radial metric component changes
	drastically from profiles like those shown in the blue lines of
	Fig.~\ref{fig:bhsols} to a profile as the one shown in red in the same figure.
	This limiting value of $\phi_0$ is well lower than the bound  one would infer
	from requiring the angular part of the metric to have a positive signature,
	$\phi_0 <  0.57$. {The reason for this behaviour is the following: for any  given $\phi_0$,
	there exists  a corresponding minimum mass $M_{min}$ that the black hole must have, in order for ensuring that 
$\phi(r)$ is real everywhere. For $\phi_ 0 \gtrsim 0.40$
	the minimum mass is larger than $M = 0.5$ (the mass value we assumed in our numerical analysis). If the value of the black hole mass 
	is less than  	 $M_{min}$,
%
%
%
%
%
the scalar field becomes imaginary at a  finite radius $r_c>0$ (depending on $M_{min}$). In this regime,
since the action and equations of motion remain real after the replacement $\phi \to i\phi$, one might  accept the possibility that
the scalar field can be imaginary, and a solution with real metric can be found for $r<r_c$. The metric components $g_{tt}$
and $g^{rr}$ match continuously to the solution for $r>r_c$:  but they and the Ricci scalar diverge at 
$r=0$.
 In order to avoid such singular
geometries, associated with imaginary scalar fields,  we must require that the mass parameter $M$ characterizing the metric components $f(r)$, $h(r)$ is larger than
 $M_{min}$. It would be interesting to find a dynamical method 
to generate such minumum mass for the system.

We can numerically plot the behaviour of the solutions for the set-up we are considering.
 The left panel of Fig.~\ref{fig:bhsols4} shows the behaviour of the scalar field solution near the minimum mass $M_{min}$ corresponding
 to $\phi_0=0.30$. For $M>M_{min}$, 
the spacetime has an event horizon, and
$\phi'$ diverges there, but the geometry and the trace of the energy-momentum tensor are regular at the horizon. For $M = M_{min}$ the spacetime is regular everywhere and does not have horizons, as shown in the right panel of Fig.~\ref{fig:bhsols4}. This is the only solution for which $\phi'_1$ is always real and vanishes precisely at $r=0$. For $M<M_{min}$, $\phi'_1$ vanishes at $r_c$. This solution can
be extended to $r<r_c$ if the scalar field is allowed to be imaginary, at the price of introducing a naked singularity in the geometry. 

	%
	\begin{figure}
		\includegraphics[width=0.49\textwidth]{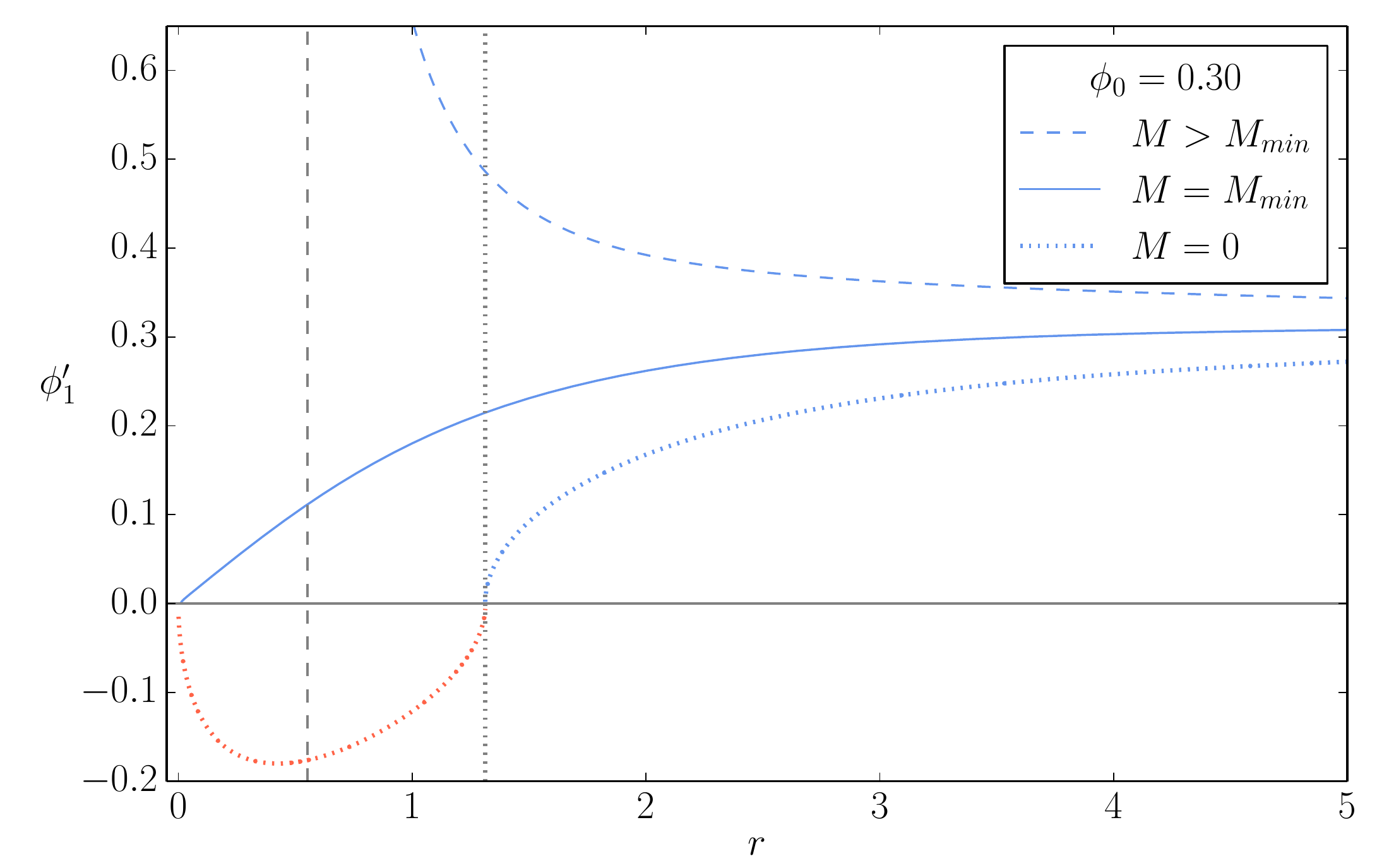} \	\includegraphics[width=0.49\textwidth]{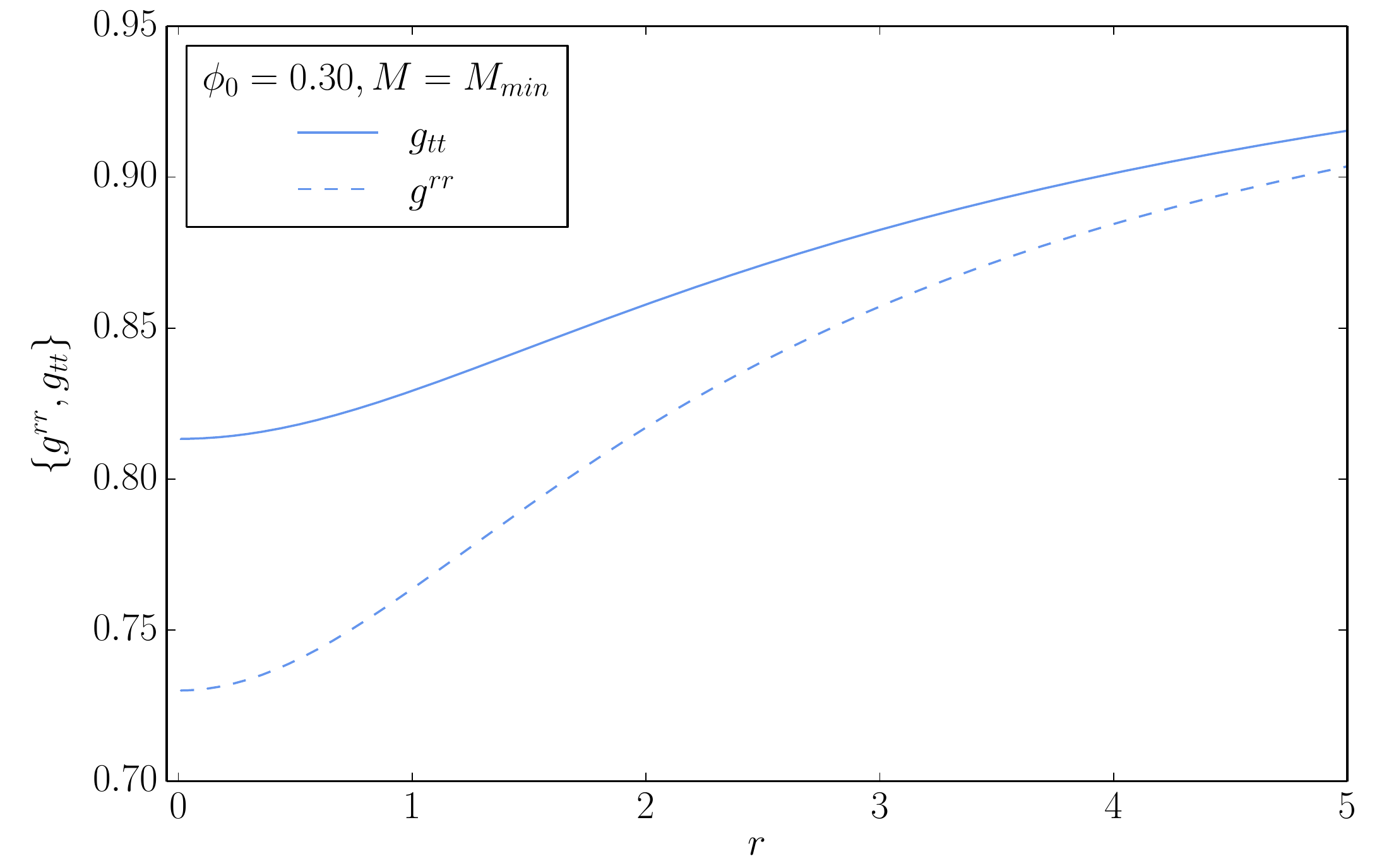} 
		\caption{\it{Vacuum geometries. The left panel shows $\phi_1'$ for a black hole geometry (dashed line) with horizon represented by the dashed vertical line, a regular geometry (solid line), and a singular geometry. 
			The radius $r_c$  where the scalar field becomes imaginary in the singular case is indicated by the dotted vertical line, and the region $r<r_c$ shows the opposite of the norm of $\phi_1'$. The right panel shows the metric fields of the regular geometry, which exists only for $M = M_{\min}$. 
				} }
		\label{fig:bhsols4}
	\end{figure}

In summary, for $M>M_{min}$ we have black hole geometries with an event horizon,
for $M=M_{min}$ we have a regular geometry without horizons, and for $M<M_{min}$ we have geometries that do not have a horizon, but they become  singular in
a region where the scalar field is imaginary: in order to have a regular geometry, we need to impose $M\ge M_{min}$. A similar situation exists in quartic Horndeski models with a
scalar field that depends only on $r$, where a minimum mass
that separates black holes from naked singularities is given in terms of the coupling constants of the model, which also determine
the (secondary) asymptotic scalar hair \cite{Babichev:2017guv}. Another analogy is the Reissner-Nordstrom black hole: given a 
electric charge $Q$, a minimum black hole mass is required to keep the singularity at $r=0$ protected by an event horizon.
}

{By repeating the analysis described above for different values of $\phi_0$, we   numerically found that the
minimum mass depends quadratically on $\phi_0$.}

	Despite the fact that the geometry does not correspond to flat space at spatial infinity,
	  the  curvature invariants go to zero, so the space-time is locally asymptotically flat, and the black hole is isolated and not affected by far away contributions to the energy momentum tensor. The asymptotic  properties of the black hole geometry seem to depend  on the scalar field
	  properties, through the deficit angle $s_0$, which at first sight enters in the computation of asymptotic charges.  On the other hand, some care is needed
	  to compute the gravitational mass through an ADM integral in theories with deficit angles. This topic has been clarified in \cite{Nucamendi:1996ac} 
	  for a geometry with the same asympotics as ours. Their work explains that the ADM energy should be properly normalized by the total angular
	  volume of the asymptotic geometry, which includes the deficit angle. Following their procedure, we find that the 
	ADM mass for our system is
	  	 \be \label{mgrav}
	 M_{ADM}\,=\,M\,
	 \ee
with $M$ the coefficient of the $1/r$ terms in the metric components $f$, $h$. Since the gravitational ADM mass is the only 
asymptotic charge for these black hole configurations, and it does not depend on the scalar parameter $\phi_0$, we conclude
that our black holes do not have  scalar hairs~\footnote{We consider `scalar hair' as any conserved
quantity which can be measured asymptotically far from the black hole, and that depends
on the scalar parameter $\phi_0$.}. This conclusion is in agreement with the recent paper \cite{Tattersall:2018map}. 
	   
	\section{Relativistic compact objects}\label{sec:kin4mat}
	In this Section we analyse non-singular, gravitationally bound star-like 
	objects with spherical symmetry, studying how  the non-minimally coupled scalar field we consider
	modifies their properties  with respect to GR configurations.
	  We find   numerical solutions that represent 
	sizeable deviations from GR solutions when the scalar parameter $\phi_0$ is large (see our
	scalar Ansatz \eqref{scal-ans1}), and that are nevertheless connected to 
	GR in the limit $\phi_0 \to 0$. Using the results of  the previous Section, we match the interior configurations for these  
	compact objects to the exterior solutions  we previously determined, in order to investigate the efficiency of Vainshtein screening
	right outside our configurations describing compact objects.

	We wish to study static, spherically symmetric configurations of matter  minimally coupled to gravity,
	\begin{equation}
	S = \int d^4x \sqrt{-g}
	  \mathcal L_c  + S_m \,, \label{ac:bhplusmatter}
	\end{equation}
	where $\mathcal L_c$ is defined in eq.~(\ref{ac:lc}), and the matter action defines the corresponding matter energy-momentum tensor as
	\begin{equation}
	T_{\mu\nu} = -\frac{2}{\sqrt{-g}}\frac{\delta S_m}{\delta g^{\mu\nu}}.
	\end{equation}
	The equations of motion for the metric result in 
	\be \label{xitmunu}
	\xi_{\mu\nu}\,=\,T_{\mu\nu}\, ,
	\ee
	where $\xi_{\mu\nu}$ is the tensor defined in eq \eqref{eq:meteq}, including metric and scalar contributions.
	We consider a perfect fluid, so that the only non-vanishing components
	of the energy-momentum tensor are
	\begin{equation}
	T^t_t = -\rho(r), \ \ T^i_j = p(r) \delta_j^i, \label{eq:tmunucomponents}
	\end{equation}
	where the Latin indices denote the spatial components of the energy-momentum tensor, and $\rho$ and $p$ characterise
	the density and pressure of the perfect fluid.
	
	 We take the metric
	Ansatz~(\ref{eq:metansatzwiths}), with 
	\begin{equation}
	 s_0  \equiv 1 - 3 \beta  \phi _0^2\,.
	\end{equation}
	In this way we guarantee that these 
	solutions are in the same coordinate frame as the exterior solutions determined in the 
	previous section. The fluid energy density and pressure can then be 
	expressed in terms of the metric components and scalar field through the relations
	\be \label{emtMETR}
	\rho(r)\,=\,-\xi_0^{\,\,0}(r)\,\hskip0.5cm, \hskip0.5cm\,p(r)\,=\,\xi_r^{\,\,r}(r)\,,
	\ee
	with $\xi_r^{\,\,r}$ the $(r,r)$ component of the tensor $\xi_\mu{}^\nu$  obtained by raising one index in equation  \eqref{eq:meteq}. 
	In order to describe the fluid
	we also need to consider an equation of state. 
	We will consider configurations of constant density,
	\be
	\rho(r)\,=\,\rho_0\,.
	\ee
	 Although it is not fully realistic, 
	this set-up allows us to obtain some  analytic results,  as well as  exact
	numerical solutions. We are interested in 
	configurations that are everywhere regular: we impose that $f'(0)\,=\,h'(0)\,=\,p'(0)\,=\,0$ to ensure regularity  at the origin of the 
	 configuration. The radial size $R_s$ of the compact object is defined as the point where the pressure profile for matter vanishes, $p(R_s)\,=\,0$.
	 
	
	Since the energy-momentum tensor is diagonal and matter is not directly 
	coupled to the scalar field, the component $\xi_{tr}$ of the metric equations of
	motion and the scalar field equation remain unchanged with respect to the vacuum case,
	 and can be solved algebraically for $\phi_1'$.  	
	In addition to the Einstein and scalar
	field equations, we impose the condition that the matter energy-momentum tensor is
	covariantly conserved,
	\footnote{This is indeed implied by the Einstein
		equations through the Bianchi Identities, given that we do not  directly couple the scalar with matter.}
	\begin{equation}
	\nabla_\mu T^{\mu}{}_{\nu} = 0.
	\end{equation}
	For an incompressible star with constant density,  the previous condition 
	gives a first order differential equation for $f(r)$ with solution ($f_0$ is  a constant)
	\begin{equation}
	f(r)\, =\, f_0  \left(p(r)+\rho_0 \right) ^{-2 }.\label{eq:fsol}
	\end{equation}
	Plugging the algebraic solution for $\phi'_1$ and \eqref{eq:fsol} into the equations of motion we
	reduce the system to two equations for $h(r)$ and $p(r)$.
	 Before entering into  this topic, it is interesting to  consider 
	 the small $r$ limit of our system, and compute the Ricci scalar. We find
\begin{equation} \label{ric-exp}
R(r\ll1)\, =\,\frac{2(1  - 3 \beta  \phi _0^2 - h_0)}{r^2}- \frac{4}{r}\frac{ {\rho_0} {h'(0)}+{h'(0)} p(0)-{h_0} p'(0)}{{\rho_0}+p(0)} + {\text{regular terms}}\, .
\end{equation}
	The coefficient of $1/r^2$ vanishes since $h_0 = s_0 =  1 - 3  \beta \phi_0^2 $, and the coefficient of $1/r$ vanishes due to
	the regularity conditions at the origin $h'(0) = 0, p'(0) = 0$. This fact distinguishes our system from the beyond Horndeski
	 set-up studied in~\cite{DeFelice:2015sya,Kase:2015gxi}, where it was shown that the angular deficit induces a singularity at $r=0$ when the scalar field depends only on $r$, due to an $1/r^2$ divergence in the Ricci scalar. In~\cite{Heisenberg:2016lux}, it was shown that this singularity can  be removed in beyond Generalised Proca theories thanks to the presence
	of a time component of the vector field. This is heuristically related to our results, since the linear dependence in $t$ of our
	scalar field can be seen as the time component of a vector field $A_\mu$ in the scalar limit $A_\mu = \nabla_\mu \phi$. 
	
	 Let us now return to discuss the solutions
	to our equations.	  We fix the constant density $\rho_0$ in the star interior, and the radius $R_s$ of the object: we would like then to determine
	solutions of our equations with the appropriate boundary conditions discussed above. 
	 In the limit of small $\beta$, we recover the standard GR solutions:
	 expanding $h(r) = h_0 (r) + \beta h_1(r) + \dots$, and 
	similarly for $p(r)$, we find that the leading terms are the GR ones corresponding to a
	TOV incompressible solution \cite{Tolman:1939jz}:
		\begin{align}
		h_0(r) & =  1-\frac{   \rho _0  r^2}{6}\, ,\\
		p_0(r) & =   \rho _0\frac{\sqrt{3 -  \rho _0 R_s^2/2}-\sqrt{3
				- \rho _0  r^2/2}}{\sqrt{3 -  \rho _0
				r^2/2}-3 \sqrt{3 -  \rho _0   R_s^2/2}}\, .
	\end{align}
	It is interesting that the GR results are recovered for small $\beta$,  
	although  we are working  in a branch of solutions that is formally disconnected from GR, and includes a non-trivial profile for the scalar field,
	\begin{equation}
	\phi_1'(r) =  \sqrt{\frac{6}{f_0}}\frac{2  \rho _0  \phi _0 \sqrt{6-R_s^2 \rho _0}}{3 \sqrt{\left(6-r^2 \rho _0\right) \left(6-R_s^2 \rho _0\right)} -6+r^2 \rho _0}\, .
	\end{equation}
	which survives in the small $\beta$ limit (analogously to the vacuum configurations, as discussed around eq \eqref{phi1ps}).
	
Outside the regime of $\beta$ small we cannot find analytical solutions, but we can attempt an approximation for low density, or investigate
the system numerically. We consider the two possibilities in what follows.

\subsection{Analytic solutions for low density}
		
%
		We assume that $h$ and $p$ can be expanded as $h(r) = h_0 + \rho_0\, h_1(r) +
		\rho_0^2 \,h_2(r)+\dots$ and $p(r) = \rho_0^2 \,p_2(r) + \rho_0^3 \,p_3(r)
		+\dots$.
		These expansions are motivated by the GR solutions for the same
		system \cite{Tolman:1939jz}. Solving the equations of motion for $h(r)$ and $p(r)$ order by order in $ \rho_0$ we find  
	\begin{align}
	h(r)  = & s_0-\frac{r^2   \rho _0}{6}+\frac{\beta \rho _0^2  \phi _0^2 s_0}{f_0 r} \left(3 r 
	-\frac{4 r \beta   s_0 }{ r^2 + 4 \beta s_0 }  - 4 \sqrt{\beta s_0}   \tan^{-1}\frac{r}{2 \sqrt{\beta s_0} } \right) \, , \label{eq:hlowdensity}\\
	p(r)  = & \frac{  \left(R_s^2-r^2\right)\rho _0^2}{24 s_0}  +\frac{\rho _0^3 }{36}  {R_s^2} \frac{R_s^2 -r^2}{4 s_0^2} \nonumber \\
	&  +\frac{\rho _0^3 }{f_0} 2 \beta ^{3/2} \phi _0^2  s_0^{1/2}\left(\frac{1}{r} \tan^{-1}\frac{r}{2 \sqrt{\beta s_0} }  - \frac{1}{R_s} \tan^{-1}\frac{ R_s }{2 \sqrt{\beta s_0} }  \right)  \, . \label{eq:plowdensity}
	\end{align}
	We remind the reader that $s_0 = 1 -3\beta\phi_0^2$. These radial profiles of the interior configuration 
	are quite different from the GR ones. 	
	 To obtain the previous solutions we impose appropriate boundary conditions
	 at the origin, and we demand a fixed radius $R_s$ for the star. We 
	 set to zero an integration constant in $h(r)$ by demanding the metric to
	be regular at the origin, and  express  the integration constant in $p(r)$
	 in terms of the radius $R_s$ where $p(r)$ vanishes. Notice
	that by requiring the star radius  $R_s$ to remain always the same as we go to higher orders
	in $\rho_0$, we   allow the central pressure to change due to the
	perturbative corrections. Up to third order, the central pressure changes to
	\begin{equation}
	p(0) = \frac{  R_s^2  \rho _0^2}{24  s_0}+ \rho _0^3 \left(\frac{   R_s^4}{ 144 s_0^2} + {\frac{\beta}{f_0} \phi _0^2} - \frac{2 \beta ^{3/2}   \phi _0^2  s_0^{1/2} }{R_s f_0} \tan^{-1}\frac{R_s }{2 \sqrt{\beta s_0} } \right)\, .
	\end{equation} 
	On the other hand, we have checked that up to third order in $ \rho_0$, the
	central value of $h(r)$ remains fixed to $h_0 = s_0$ due to non-trivial
	cancellations between the higher order corrections. (This ensures that the Ricci scalar $R$ remains regular at the origin, see eq \eqref{ric-exp}). 
	
	{The limit of empty object,  $\rho_0 \to 0 $ in the solutions for $h$ and $p$ shown in eqs.~(\ref{eq:hlowdensity}, \ref{eq:plowdensity}) has to be taken with some care. These profiles solve the equations of motion obtained after imposing the covariant conservation of
	matter, eq.~(\ref{eq:fsol}), and are not necessarily continuously connected to the vacuum solutions~(\ref{eqs:asympkinsolf}-\ref{eqs:asympkinsolphi}). When $\rho_0 = 0$ the pressure vanishes as well, and the continuity equations loses its physical interpretation. However, to be consistent with the system of equations that we solved,  we have to require 
	$f_0 \sim \rho_0^2$, so that  $f$ acquires a finite value. 
	On the other hand, the vacuum solutions do not admit in general a constant
	profile for $f$: the only way to make  this possible is to impose that  $M$ and $\phi_0$ vanish. Thus, if we want that the limit 
	$\rho \to 0$ is continuously connected to a solution of the vacuum equations of motion, 
	the continuity equation imposes that $\phi_0 = 0$ when $\rho_0 = 0$, and the solution reduces to Minkowski spacetime with 
	a constant scalar field. }
	\subsection{Numerical solutions}\label{comnumse}
	
	We now  investigate interior configurations using numerical methods, in a regime
	where $\beta$, $\phi_0$ and $\rho_0$ are not necessarily small. As we 
	shall learn, we find interesting conditions on the parameters involved in order to get regular solutions, 
	which can indicate new ways to constrain the scalar-tensor theories under consideration. Our analysis  will
	focus to study the compactness of the stellar object, a physical quantity that will be helpful to point out differences
	with GR results. 
	
	\smallskip
	
We compute interior solutions for different values of $\phi_0$
	and $\rho_0$ by solving numerically the system of
	equations derived from~(\ref{ac:bhplusmatter})-(\ref{eq:tmunucomponents}), with the metric
	Ansatz~(\ref{eq:metansatzwiths}) and $s_0 = 1 - 3 \beta \phi_0^2$. The initial conditions
	are set at  small radius, and are determined by 
	Taylor expanding the equations of motion around $r=0$, imposing that
	at the origin the fields behave as $h(0) = 1 - 3 \phi_0^2 \beta$, $h'(0)
	= 0$, and $p'(0) = 0$,  and solving for $h''(0)$ and $p''(0)$.  
	We work in units where $M_{Pl}=1$, and we fix for definitiness  $\beta = 1$; we work 
	imposing a fixed radius $R_s$ for the star,  {$R_s =
		1.5 $}.
	
	The parameters that need to be provided to the system of equations in order to fully
	determine the radial metric component $h(r)$, the pressure $p(r)$, and their
	derivatives near the origin are the constant density $\rho_0$, the value of
	$\phi_0$, and the central pressure $p(r=0)\,=\,p_0$ -- which controls the radius $R_s$ of
	the resulting configuration. To explore this parameter space we
	choose  arbitrary values of $\rho_0$ and $\phi_0$, and we select $p_0$ by
	requiring that the resulting configurations have a given radius (that we choose arbitrarily).
	 In GR, the central pressure that satisfies this requirement can be computed exactly for stars of radius $R_s$,  by
	evaluating the TOV incompressible solution for the pressure at the origin \cite{Tolman:1939jz}:
	\begin{equation}
	p_{0,GR} = \rho_0 \frac{\sqrt{3  - R_s^2  \rho_0/2} - \sqrt{3
		}}{\sqrt{3 } - 3 \sqrt{3  - R_s^2  \rho_0/2}}\, .
		\label{eq:p0gr}
		\end{equation} 
	For our beyond Horndeski system we do not have an analytic method for determining the
	central pressure associated with  a configuration with a given radius $R_s$. Thus,
	we proceed numerically  by fixing $\rho_0$ and $\phi_0$ and shooting $p_0$ until we find a
	solution with the desired radius. For any $\rho_0$ and small $\phi_0$,
	Eq.~(\ref{eq:p0gr})  serves as seed  for  $p_0$: then the resulting $p_0$ serves
	respectively as seed for the central pressure of a configuration with a higher
	$\phi_0$, and this process is repeated until $\phi_0 \sim 0.5$, where we approach
	the limit imposed by requiring that the sign of the angular component of the 
	metric is preserved.

	\begin{figure}
		\includegraphics[width=0.8\textwidth]{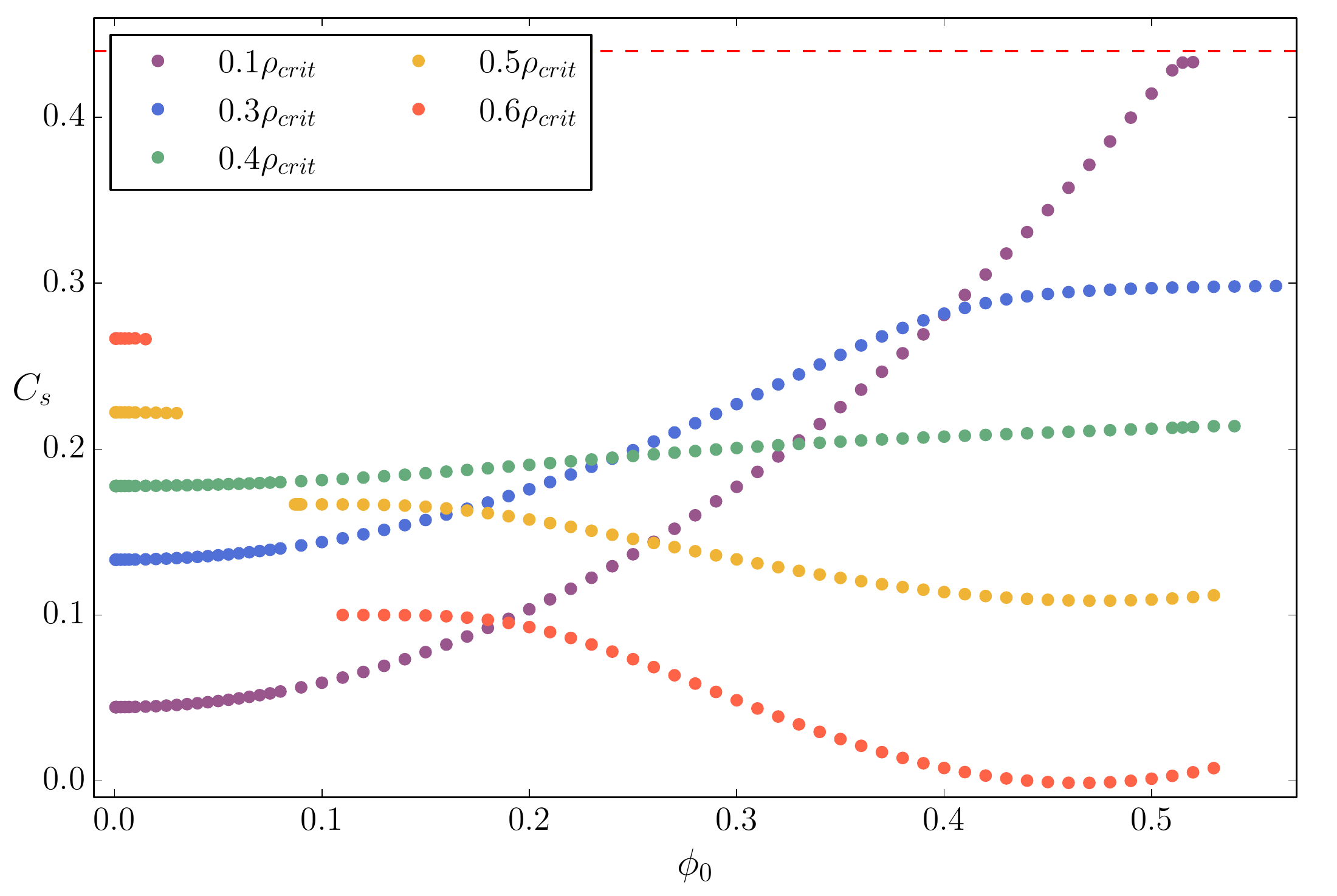}
		\caption{\it Compactness of constant density objects for  $\beta = 1 $
			and different values of $\phi_0$. The dashed line
			shows the GR limit for the compactness, which is obtained only for
			an object with the critical density $\rho_{crit}$. The density of each solution is indicated by the point color. The gap in the sequence of solutions with $\rho_0 = 0.5 \rho_{crit}$ and $\rho_0 = 0.6 \rho_{crit}$ is an effect of the scalar field contributions, as explained in the main text.} 
		\label{fig:starcompactness}
	\end{figure}

	We fix the radius of the star at a value $R_s = 1.5$, hence it is convenient to parametrise the density
	$\rho_0$ in terms of
	the critical density in GR for an object of a given  $R_s$. We do so by writing
	$\rho_0 = A \rho_{crit}$, where $A$ is a constant in the range $(0,1)$ and $\rho_{crit}$
	is the critical density of a compact object of constant density in GR (see, e.g., \cite{Carroll:2004st}),
	\begin{equation}	\label{drhoc}
	\rho_{crit} =  \frac{16 }{3 R_s^2  } .
	\end{equation} 
	Solutions with $\rho_0 \geq \rho_{crit}$ do not exist in GR, and we do not find evidence of their existence 
	in the beyond Horndeski model under consideration.
	
	\begin{figure}
		\includegraphics[height=6cm]{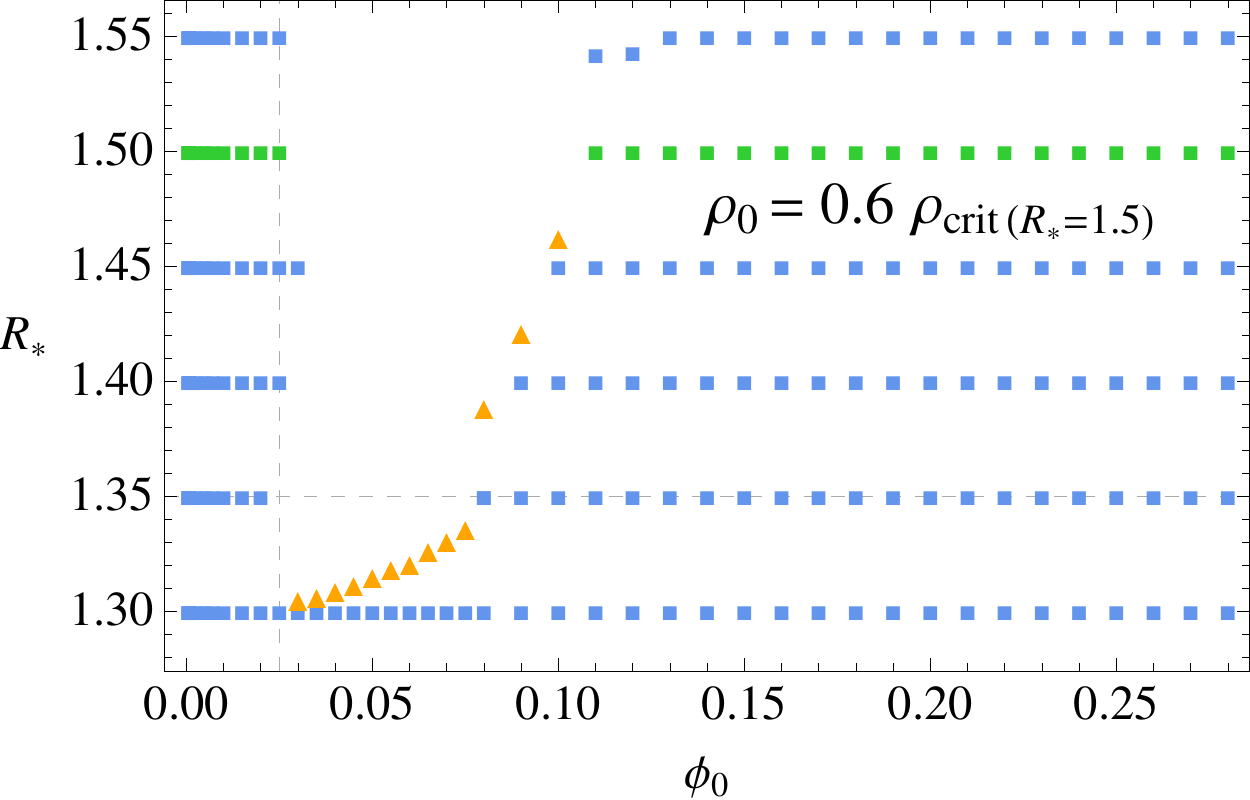}
		\caption{\it $R_s$-$\phi_0$ parameter space for a fixed $\rho_0$, equal to 60$\%$ of the critical density
			of an object of radius $R_s = 1.5$. Green points correspond to the solutions shown in red in Fig.~\ref{fig:starcompactness}. For $R_s \gtrsim 1.3$ there is a region of the parameter space where
			we do not find solutions. The points  coloured in orange are referred to in the next Figure. } 
		\label{fig:smallerradius}
	\end{figure}

	\smallskip
	We apply the results of this numerical method to investigate a physically relevant quantity, the stellar compactness,  which allows us 
	to find constraints on the parameters involved, and also to point out differences with GR configurations.
	The intrinsic stellar compactness,  which we plot in Fig.~\ref{fig:starcompactness}, is defined as
	\begin{equation}\label{intcom}
		C_s = \frac{m(R_s)}{R_s}.
   \end{equation}
   In the previous expression, the mass of the star 
	  $m(R_s)$ corresponds  the value at $R_s$ of the the mass function $m(r)$ defined by expressing the metric component $h(r)$ in the stellar
	  interior  as
		\begin{equation}
		h(r) = 1 - \frac{2m(r)}{r}\, ,
		\end{equation}
that is, including within $m(r)$ all the radial dependence of corrections to the Schwarzschild metric due to matter and scalar field. The compactness
defined in this way only includes contributions of the interior of the star -- this is why we call it intrinsic --
 and it is in principle different from the compactness as measured
by an asymptotic observer, which we shall  discuss in the next subsection. 
	Such difference is important for characterizing the efficiency of the screening mechanism in proximity of the object surface. 
	
	Each point in Fig.~\ref{fig:starcompactness} represents a configuration of matter with
	radius $R_s = 1.5$,  density indicated by the point colour, $\phi_0$ by the $x$-axis, and stellar compactness by the $y$-axis. We observe the following properties:
	\begin{itemize}
	\item High stellar compactness is possible for configurations with a low density of matter: this is  due to the large  contributions of the
	scalar profile for characterizing the internal geometry of the system.  
	\item 	For $\rho_0 \gtrsim 0.4 \rho_c$, there exists a range of values of $\phi_0$ where we
		       cannot find configurations with the desired radius $R_s$. The reasons for this will be  explored in some length below.
    \item The intrinsic stellar compactness 
              does not exceed the GR limit $C = 4/9 \approx 0.44$ (see, e.g., \cite{Carroll:2004st}). 
              This is in contrast to what
              happens in vector-tensor theories \cite{Chagoya:2017fyl}, but similar findings have been reported for
              a subset of Horndeski gravity \cite{Maselli:2016gxk}.
               In the next section we show that this is true even when the effects from the exterior solution are taken into account.  
	\end{itemize}
	
	The fact that we  find a gap in the range of allowed stellar densities is interesting, and deserves some more words since it can 
	suggest ways to test and constrain the parameter space of relativistic compact objects in scalar-tensor theories.  We 
	 	 investigate  in more detail what happens in the region where we cannot find solutions with $R_s = 1.5$. We fix the density to be $60\%$ of the critical density: the green points in
	Fig.~\ref{fig:smallerradius} correspond to the configurations shown in red in Fig.~\ref{fig:starcompactness}.
	From $\phi_0 \approx 0.02$ to $\phi_0 \approx 0.10$ we do not find solutions with $R_s = 1.5$. Indeed, around $\phi_0 \approx 0.02$ there is a drastic change in the maximum radius, which falls 
	to about $R_s = 1.3$, as shown by the orange points in Fig.~\ref{fig:smallerradius}. The blue points in the
	same figure show configurations with the same density as the green and orange points, but for different radius: these are drawn in order to outline the region where solutions do not exist. 

	\begin{figure}
		\includegraphics[height=4.9cm]{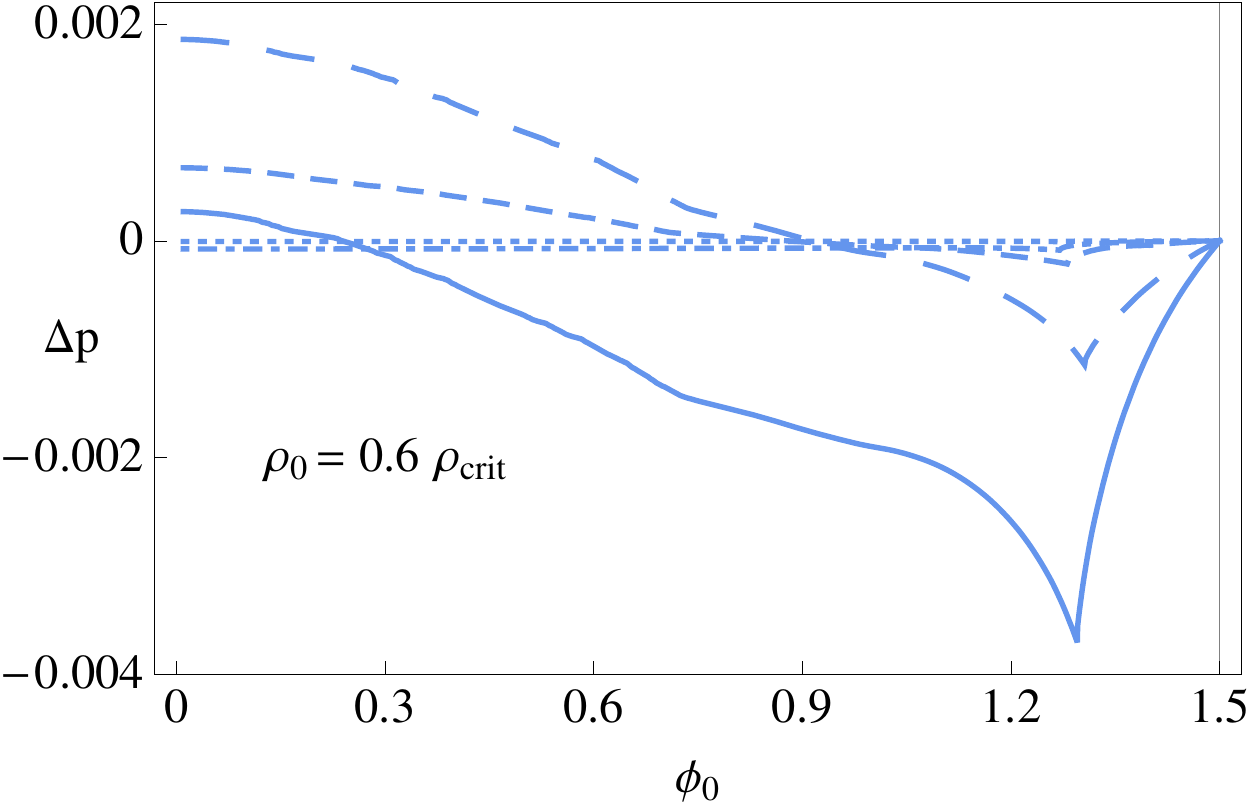} \ \ 
		\includegraphics[height=4.9cm]{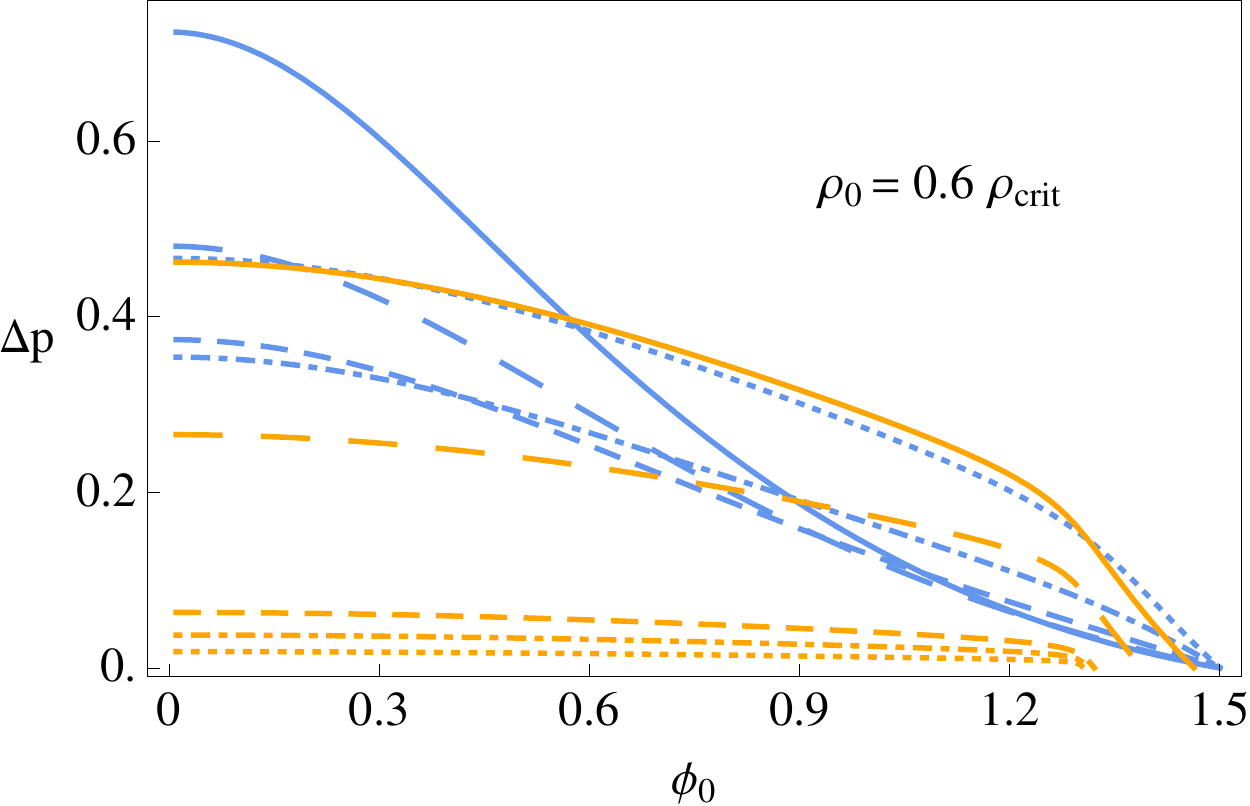}
		\caption{\it Difference between the pressure in GR and in the beyond  Horndeski model under consideration for configurations with $60\%$ of the critical density in  GR for an object of radius $R_s =1.5$. The left panel shows $\Delta p$ for solutions before the gap along the sequence of green points in Fig.~\ref{fig:smallerradius}, while the right panel 
			corresponds to solutions in the gap (orange curves correspond to orange points in Fig.~\ref{fig:smallerradius}) and along the sequence of green points after the gap.
		 } 
		\label{fig:deltapressure}
	\end{figure}

	\smallskip
	
		The origin of an interval in the parameter space where solutions do not exist, and in particular of the drastic	change of $R_s$ near the lower end of this interval in $\phi_0$, can be understood with the help of Fig.~\ref{fig:deltapressure}, where we plot a quantity defined as  
%
		\begin{eqnarray}
		\Delta  p(r) &=& p(r) - p_{GR}(r)\,, \nonumber
		\\&=&\xi_r^{\,\,r}-G_r^{\,\,r}\,,
		\end{eqnarray}
		where $\xi_r^{\,\,r}$ is the $(r,r)$ component of the left-hand-side of the equations of motion for the metric (see eqs \eqref{xitmunu} and  \eqref{emtMETR}),
		while $G_r^{\,r}$ the $(r,r)$ component of the Einstein tensor calculated on the configuration we examine, with no contributions from the scalar field (recall we work in units $M_{Pl}=1$). 
		The quantity $\Delta  p(r)$ describes the specific contributions to the total pressure which can be associated with the scalar field. 
%
%
%
%
%
		The left panel shows $\Delta p$ for solutions with $\phi_0 < 0.02$ and $R_s=1.5$; the solid line corresponds to the last configuration along the sequence of green points before the gap in Fig.~\ref{fig:smallerradius}.  We see that $\Delta p$ has a minimum at some radius significantly smaller than $R_s$. Based on this, we speculate that for
		$\phi_0 \gtrsim 0.02$, $\Delta p$ acquires large negative values, whose size 
		 is sufficient
		  to drive the total pressure $p(r)$ to zero
		at a radius smaller than the value $R_s\,=\,1.5$, that we  initially  fix by means of the initial conditions. Since the star radius is defined as the point where the pressure vanishes, the
		 large scalar contribution   to the pressure makes the radius smaller than 
		 the one we impose. 
		 Hence, we learn that there are regions in the parameter space of the scalar-tensor theory under consideration where -- due to large contributions
		 associated with  the scalar field --  there do not  exist 
		 compact configurations for certain radii and energy densities.

		 	As mentioned above, 
	we can overcome the problem and find solutions by 
	changing some of the  conditions, for example  by reducing the stellar size $R_s$.  The physical requirement is that the total
	 pressure 
	  vanishes at $R_s$. In order to find  
	  the correct value of $R_s$ where this happens  maintaining the same energy density $\rho_0$,
	   we thus need to change 
	the central pressure to a smaller value such that both the GR and scalar field contributions to the pressure vanish at the same point. The configurations with maximum radius that we  find are shown with orange markers in 
	Fig.~\ref{fig:smallerradius}, and the profiles of $\Delta p$ associated to them are shown with orange lines
	in the right panel Fig.~\ref{fig:deltapressure}. The curves shown in blue in the same plot instead correspond 
	to configurations along the sequence of green points, to the right of the gap.


These results show  that the scalar-tensor theory under consideration imposes more stringent constraints on the stellar properties with respect 
to GR, since we identified forbidden regions on the the energy density-radius plane, which depend on the value of        
$\phi_0$, and
where regular star configurations do not exist. 
  In a more refined version of our analysis,  considering a polytropic equation of state,  this fact can suggest observable tests for
  the parameter space of 
   these scalar-tensor theories, 	
		 which would be excluded in case compact objects are found within the forbidden regions.

	\subsection{Matching of interior and exterior solutions}
	
\begin{figure}
		\includegraphics[height=6cm]{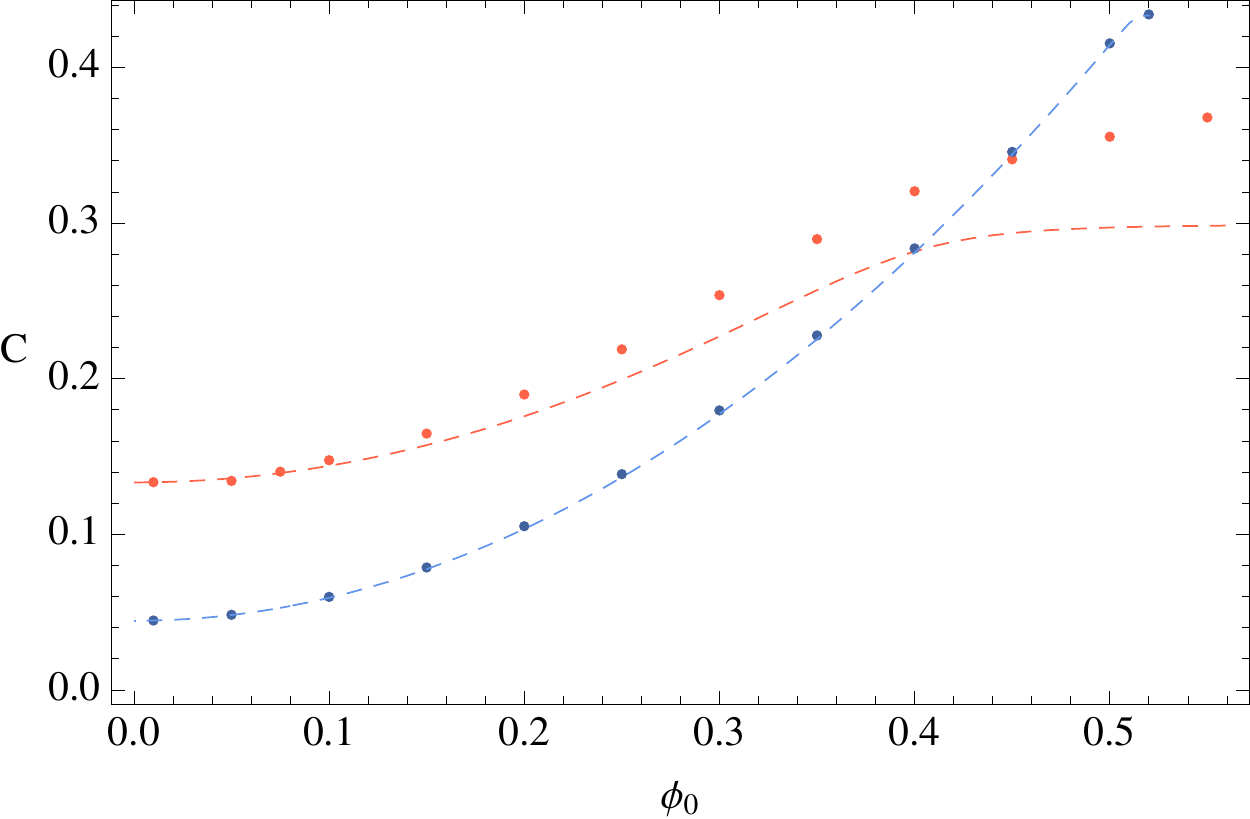}
		\caption{\it Compactness of constant density configurations for  $\beta = 1 $. The case $\rho_0 = 0.1 \rho_{crit}$ is shown in blue, and $\rho_0 = 0.3 \rho_{crit}$ in red. The dashed lines show the stellar compactness,
			and the points show the compactness measured asymptotically. The scalar field can
			have a relevant impact in the asymptotic compactness, but not enough as to
			get values of $C$ higher than the GR limit $C\approx 0.44$.} 
		\label{fig:starcompactnesscomparisonintext}
	\end{figure}

	In section~\ref{sec:bhsol} we learned that static, spherically symmetric vacuum
	solutions to the equations of motion derived from~(\ref{ac:lc}) do not correspond  exactly to GR configurations in
	vacuum, since they  differ
	from the Schwarzschild solution by an amount controlled by $\phi_0$ and $\beta$.
	 Therefore, the extrinsic compactness measured  by an observer far away from
	 a compact object can be different from the intrinsic quantity we studied in Section \ref{comnumse}, 
	--  eq \eqref{intcom} and below -- due to contributions from the exterior part of the geometry.
	 To investigate  how
	large these contributions are, we take the  very same values of the metric and scalar field
	at a position $R_s$ from the solutions shown found in  Section  \ref{comnumse}, and we use these
	values as initial conditions to integrate numerically the vacuum equations from
	$R_s$ outwards. At large $r$, we compute the gravitational  mass using the
	asymptotic solutions~\eqref{eqs:asympkinsolf}-\eqref{eqs:asympkinsolh}.
	The matching of the interior and exterior solutions at $R_s$ is
	straightforward, and  we match $\phi'$, $g_{rr}$, $g'_{rr}$ and $g_{tt}$ at that point.

	In Fig.~\ref{fig:starcompactnesscomparisonintext} we 	reproduce the intrinsic  stellar compactness of configurations
	with $\rho_0 = 0.1 \rho_{crit}$ and $\rho_0 = 0.3 \rho_{crit}$ (dashed lines), and we show the asymptotic, extrinsic 
	compactness for some of these configurations (points). Interestingly, even when the effects
	from the exterior solution are taken into account, the compactness does not exceed the GR limit
	$C = 4/9$. Also, notice that for low density the screening of the exterior solution is highly efficient: even though the 
	scalar field introduces large modifications to the stellar compactness, the scalar contributions in the exterior 
	are negligible, and extrinsic and intrinsic values of the compactness almost coincide. On the other hand, for higher values of 
	the stellar energy density, the values of the extrinsic and intrinsic compactness differ for large values of the parameter $\phi_0$. This implies that the effect
	of the scalar field in this regime is relevant also outside the object, and not only on its interior. 
	
	\section{Discussion}
	
	The  recent observation of gravitational waves from a neutron star merger 
	 GW170817 and its associated electromagnetic counterpart GRB170817A has changed 	 our perspective on scalar-tensor
	 theories. One possibility is to focus only on the simplest theories where the graviton speed $c_{GW}$ is automatically equal to one;
	 the other is to consider richer systems where   this condition  is obtained at the price of tuning some parameters. In this work we considered
	 the second possibility, studying the physics of compact objects in a theory of beyond Horndeski with  $c_{GW}=1$ that includes the scalar
	 kinetic term.

		We focussed on black hole and relativistic star
		configurations which are locally asymptotically flat, that
		can be    continuously connected  to  GR configurations, and that have been 
		less explored in the literature.  Depending on a parameter controlling the scalar field, $\phi_0$, our solutions can be very similar to 
		GR when $\phi_0$ is small, while they can provide sizeable corrections to it when  $\phi_0$ is larger. 
This shows that a Vainshtein screening mechanism, which is very effective to reproduce GR predictions in a weak gravity limit, can 
		be less so in strong gravity regimes. 
		
		For what respect black hole configurations, we shown that our geometries are  characterized asymptotically by an angular deficit, due to presence
		of the scalar kinetic term, and are
		equipped with regular horizons provided that the black hole mass is larger than a value depending on the scalar parameter $\phi_0$.
		Our geometries have not scalar hairs, despite the fact that the scalar has a profile that extends asymptotically far from the black hole. 
		 The black hole solutions can be  more compact
	than the Schwarzschild black hole, thanks to the effect of the scalar field. 
The angular  deficit could be detected 
	by its effect on geodesics and light propagation \cite{Barriola:1989hx,Shi:2009nz}.

		We also studied regular relativistic compact objects corresponding to incompressible stars with constant energy density. The scalar field
		modifies  properties of the star  
		as its compactness, allowing for stars that are twice as compact as neutron stars with the same matter density. These deviations 
		from GR can be accessed observationally, for example through
	quantities that depend on the tidal deformability of a star, which is directly affected by the compactness \cite{Hinderer:2007mb,Paschalidis:2017qmb}.
		We also found that there are forbidden regions in parameter space where regular star configurations of given radius and energy density cannot be found, depending
		 of the scalar field profile. In a more refined version of our analysis,  considering a polytropic equation of state,  this fact can suggest observable tests for 
		 the parameter space of these scalar-tensor theories, 	
		 which would be excluded in case objects are found in the forbidden regions.

	By analysing the difference between our interior and exterior solutions and their GR counterparts, we  
	numerically investigated  the efficiency  of the screening of the scalar field inside and outside the relativistic star. We found that including the standard
	kinetic term of the scalar field breaks the perfect screening of vacuum solutions, not only because of the angular deficit but 
	also because the time and radial components of the metric acquire corrections that distinguish  them from the Schwarzschild
	solution in the exterior of the object. Nevertheless, there are situations where such deviations from a Schwarzschild solution are small
	in the exterior, while the corrections to the interior metric
	are large with respect to GR. We cannot find the opposite situation -- corrections that are large in the exterior but small in the interior. This
	indicates that the breaking of screening is more severe in the interior solutions.
	
	Much work is left for the future. It is interesting to continue to investigate the physics of compact objects in other scalar-tensor theories with $c_{GW}=1$, for realistic equations of state for the star interior.

		    \acknowledgments
    We are partially supported  by the STFC grant ST/P00055X/1.
	
		\begin{appendix}
			\section{Equations of motion} \label{app:a}
			The covariant equations of motion derived from action~(\ref{ac:lc}) with $G_4 = M_{Pl}^2 + {M_{Pl}^{-2}\beta} \, X$ are
			\begin{align}
				0  = & \mpl^{-2}\beta \left[ \phantom{\frac{}{}}[\Phi] R  + \nabla_\alpha R \nabla^\alpha
				\phi  + \frac{ R_{\alpha\beta}\langle \Phi \rangle - [\Phi]^2\langle \Phi
					\rangle  + \langle \Phi \rangle [\Phi^2] + 2 [\Phi] \langle \Phi^2 \rangle - 2
					\langle \Phi^3 \rangle }{ X^2}   \right.\nonumber \\
				& \left. - \frac{[\Phi]^3 - 3 [\Phi] R_{\alpha\beta}\phi^\alpha\phi^\beta -3
					[\Phi] [\Phi^2] -  \phi^\alpha \phi^\beta \phi^\sigma \nabla_\sigma
					R_{\alpha\beta} + 2 [\Phi^3] + 2 R_{\alpha\sigma\beta\delta} \phi^\alpha
					\phi^\beta  \Phi^{\delta\sigma}}{X} \right]+ [\Phi] , \label{eq:esceq}\\
				0 = & \left(\mpl^2 +\frac{\beta X}{\mpl^2} \right)G_{\mu\nu} -\frac{g_{\mu\nu} X}{2}
				-\frac{ \phi_\mu  \phi_\nu}{2}  - \frac{\mpl^{-2} \beta}{4 X^2}\left( 2 g_{\mu\nu} \langle  \Phi\rangle^2  -
				4\phi^\alpha \langle \Phi \rangle \Phi_{(\mu|\alpha}  \phi_{|\nu)} + 2 \langle
				\Phi^2 \rangle \phi_\mu \phi_\nu \right) \nonumber \\
				& -\frac{\beta }{\mpl^2}  \left[ g_{\mu\nu} (\nabla_\alpha[\Phi]
				\phi^\alpha  +   R_{\alpha\beta}\phi^\alpha \phi^\beta + [\Phi^2] ) -
				\nabla_\alpha  \Phi_{\mu\nu} \phi^\alpha - \Phi_{\nu\alpha}\Phi^\alpha_\mu
				-R_{\mu\alpha\nu\beta}\phi^\alpha \phi^\beta   + \frac{R \phi_\mu \phi_\nu}{2}  \right]  \nonumber \\
				&  -\frac{ \beta \mpl^{-2}}{2X}\left[
				g_{\mu\nu} (3 \langle \Phi^2 \rangle +  \phi^\alpha \phi^\beta
				\nabla_\sigma \Phi_{\alpha\beta} \phi^\sigma )-2 \phi^\alpha \phi^\beta (
				\Phi_{\mu\alpha}\Phi_{\nu\beta}  +  \nabla_\beta
				\Phi_{(\mu|\alpha}\phi_{|\nu)} )+ 2 [\Phi] \phi^\alpha
				\Phi_{(\mu|\alpha}  \phi_{|\nu)}    \right. \nonumber \\
				& \left.  - 4 \Phi_{\beta\alpha} \phi^\alpha
				\Phi^\beta{}_{(\mu} \phi_{\nu)} + \phi_\mu \phi_\nu( 2
				R_{\alpha\beta}\phi^\alpha \phi^\beta   + \nabla_\alpha [\Phi] \phi^\alpha  + 2
				[\Phi^2] - [\Phi]^2 )  \right]    
				\\
				\equiv& \,\,\xi_{\mu\nu} \, .
				\label{eq:meteq}
			\end{align}
			Under the spherically symmetric ansatz~(\ref{eq:metansatzwiths}), the $(r,r)$, $(t,t)$ and $(t,r)$ components of the equations of motion become
			\begin{align}
		0 =&	\left(f h \phi _1'{}^2 -\phi _0^2\right) \left(-4 f s_0+r^2 \phi _0^2+f \left(4 h+4 r h'+h r^2 \phi _1'{}^2\right)\right) +2 \beta  \left\{\phi _0^4 \left(h-s_0+r h'\right) \right. \nonumber \\ 
		& \left. +2 f h \phi _0^2 \phi _1' \left[\left(2 h+3 r h'\right) \phi _1'+4 h r \phi _1''\right]
		- f^2 h^2 \phi _1'{}^3 \left[\left(h-s_0+3 r h'\right) \phi _1'+4 h r \phi _1''\right]\right\} \, , \label{eq:ssstt} \\
			0 = & f \left(\phi _0^2-f h \phi _1'{}^2\right) \left(4 f s_0+r^2 \phi _0^2+h \left(-4 f-4 r f'+f r^2 \phi _1'{}^2\right)\right) +2 \beta  \left\{\phi _0^4 \left(f s_0 -f h + h r f'\right) \right. \nonumber \\ 
			& \left. -f^2 h^2 \left[f s_0+3 h \left(f+r f'\right)\right] \phi _1'{}^4+2 f h^2 r \phi _0^2 \phi _1' \left(3 f' \phi _1'+2 f \phi _1''\right)\right\}  \, , \label{eq:sssrr}  \\
		0 = & \phi _0 \phi _1' \left\{f r^2 \left(f h \phi _1'{}^2 - \phi _0^2\right)+2 \beta  \left[\phi _0^2 \left(f h-f s_0-2 h r f'+f r h'\right)+f h \left(f s_0+h f+r h f'\right) \phi _1'{}^2\right]\right\} \, . \label{eq:ssstr} 
			\end{align}
			The scalar equation is automatically satisfied by the algebraic solution to eq.~(\ref{eq:ssstr}) for $\phi'$. 
		\end{appendix}

\end{document}